\documentclass[preprint,12pt,nofootinbib]{revtex4}

\usepackage[]{graphicx}\graphicspath{{graph/}}
\usepackage{amsmath,amssymb,amsfonts}
\usepackage{epstopdf}
\usepackage{hyperref}
\hypersetup{
  colorlinks,
  citecolor=magenta,
  linkcolor=blue}

\renewcommand{\Re}{{\rm Re\,}}

\newcommand{\p}{{\partial}}

\newcommand{\intinf}{\int_{-\infty}^\infty}
\makeatletter
\newcommand{\lambdabarb}{{\mathchoice
  {\smash@bar\textfont\displaystyle{0.25}{1.2}\lambda}
  {\smash@bar\textfont\textstyle{0.25}{1.2}\lambda}
  {\smash@bar\scriptfont\scriptstyle{0.25}{1.2}\lambda}
  {\smash@bar\scriptscriptfont\scriptscriptstyle{0.25}{1.2}\lambda}
}}
\newcommand{\smash@bar}[4]{%
  \smash{\rlap{\raisebox{-#3\fontdimen5#10}{$\m@th#2\mkern#4mu\mathchar'26$}}}%
}
\makeatother
\begin{document}


\title{Cooling Rate for Microbunched Electron Cooling without Amplification
 }

\author{G. Stupakov}
\affiliation{SLAC National Accelerator Laboratory, Menlo Park, CA 94025}

\begin{center}
\end{center}

\begin{abstract}

The Microbunched Electron Cooling (MBEC) proposed by D. Ratner is a promising cooling technique that can find applications in future hadron and electron-ion colliders. In this paper, we 
develop a new framework for the study of MBEC which is based on the analysis of the dynamics of microscopic 1D fluctuations in the electron and hadron beams during their interaction and propagation through the system. Within this framework, we derive an analytical formula for the longitudinal cooling rate and benchmark it against 1D computer simulations. We then calculate the expecting cooling time for a set of  parameters of the proposed electron-ion collider eRHIC in a simple cooling system with one chicane in the electron channel. While the cooling rate in this system turns out to be insufficient to counteract the intra-beam scattering in the proton beam, we discuss how the electron signal can be amplified by two orders of magnitude through the use of plasma effects in the beam.

\vfill
%
\end{abstract}

\maketitle

%
\section{Introduction}\label{sec:1}
%

\begin{figure}[htb]
\centering
\includegraphics[width=0.7\textwidth, trim=0mm 0mm 0mm 0mm, clip]{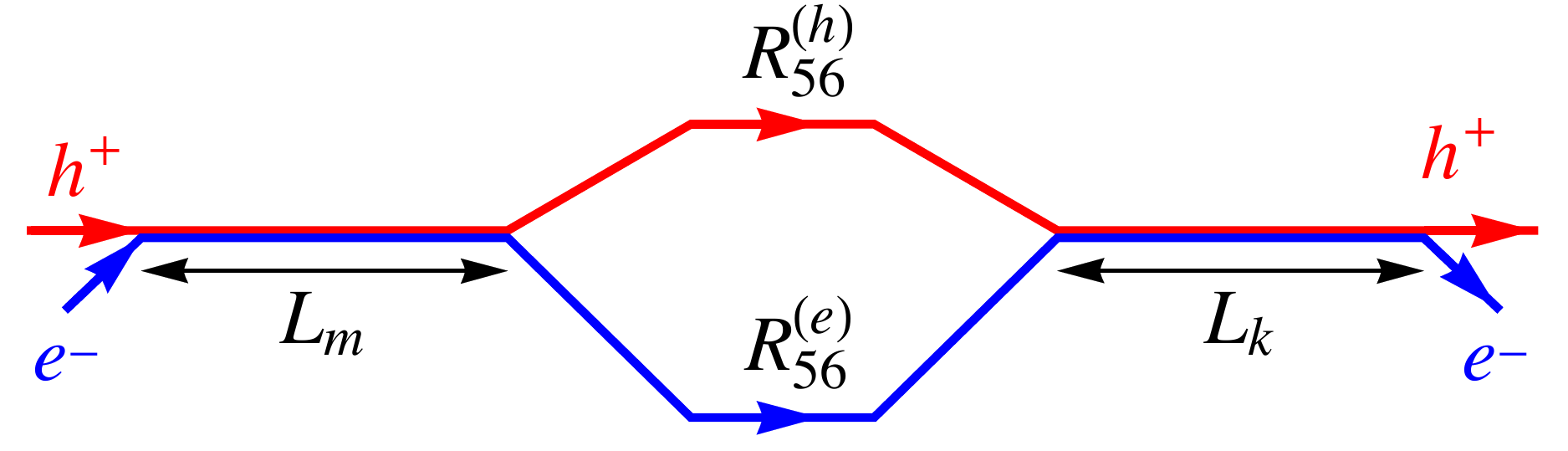}
\caption{Schematic of the microbunched electron cooling system. Blue lines show the path of the electron beam, and the red lines indicate the trajectory of the hadron beam.}
\label{fig:1}
\end{figure}

The idea of coherent electron cooling has been originally proposed by Ya. Derbenev~\cite{Derbenev:1991ur} as a way to achieve cooling rates higher than those provided by the traditional electron cooling technique~\cite{MESHKOV1997459,nagaitsev06etal}. The mechanism of the coherent cooling can be understood in a simple setup shown in Fig.~\ref{fig:1}. 
An electron beam with the same relativistic $\gamma$-factor as the hadron beam, co-propagates with the hadrons in a section of length $L_m$ called the ``modulator''. In this section, the hadrons imprint microscopic energy perturbations onto the electrons via the Coulomb interaction. After the modulation, the electron beam passes through a dispersive chicane section, $R_{56}^{(e)}$, where the energy modulation of the electrons is transformed into a density fluctuation referred to as ``microbunching''\footnote{In a long modulator section the microbunching can be generated directly in the modulator when the energy modulation is converted into a density fluctuation through plasma oscillations~\cite{PhysRevE.78.026413}.}. Meanwhile, the hadron beam passes through its dispersive section, $R_{56}^{(h)}$, in which more energetic particles move in the forward direction with respect to their original positions in the beam, while the less energetic trail behind. When the beams are combined again in a section of length $L_k$ called the ``kicker'', the electric field of the induced density fluctuations in the electron beam acts back on the hadrons. With a proper choice of the chicane strengths, the energy change of the hadrons in the kicker leads, over many passages through the cooling section, to a gradual decrease of the energy spread of the hadron beam. The transverse cooling is achieved in the same scheme by introducing  dispersion in the kicker for the hadron beam.

In most cases, the cooling rate in the simple setup shown in Fig.~\ref{fig:1} is not fast enough for practical applications. It can be considerably increased if the fluctuations in the electron beam are amplified on the way from the modulator to the kicker. Litvinenko and Derbenev proposed to use for this purpose the gain mechanism of the free electron laser (FEL)~\cite{litvinenko09}. While this may be sufficient for some applications, one of the drawbacks of this approach is a narrow-band nature of the FEL amplifier that may not provide enough gain before the amplified signal saturates~\cite{stupakov13cec}. Following an earlier study by Schneidmiller and Yurkov~\cite{schneidmiller_2010} of microbunching dynamics for generation of coherent radiation, Ratner proposed a broadband amplification mechanism~\cite{Ratner2013} in which the amplification is achieved through a sequence of drifts and chicanes such that the density perturbations in the drifts execute a quarter-wavelength plasma oscillation. In a recent paper~\cite{2018arXiv180208677L}, Litvinenko and co-authors put forward an idea to use a parametric instability in the electron beam when the transverse size of the beam is periodically varied when it propagates through the cooling system.

A considerable effort has been devoted to theoretical and computational analysis of various aspects of coherent cooling~\cite{PhysRevE.78.026413,PhysRevSTAB.16.124001,Ratner2013,PhysRevSTAB.18.044001,PhysRevSTAB.17.101004}. However, to our knowledge, the theory is still lacking a simple formula that would allow to predict the cooling rate and its scaling with the main parameters of the cooling system, similar to simple formulas available for the traditional electron cooling~\cite{nagaitsev_e_cool}. In this paper, we derive such a formula, Eq.~\eqref{eq:67}, for a system shown in Fig.~\ref{fig:1}. While this system, as we will see below, may not provide the required cooling rate for some applications, we believe that its study constitutes a necessary first step toward a more complex design which uses amplification stages to increase the cooling rate. We plan to carry out a quantitative analysis of MBEC with amplification cascades in a separate work.

The original approach in Ref.~\cite{Ratner2013} was based on the analysis of hadron-to-hadron interactions in the cooling system as hadrons co-propagate with the electrons through the drift sections and pass through the chicanes. In this approach, the self-interaction of hadrons, under certain conditions, can lead to the cooling while the interaction of different particles causes the energy diffusion in the beam. Conceptually a similar treatment is used in the classical stochastic cooling~\cite{BisognanoLeemann1982} where the BBGKY equations~\cite{liboff1969introduction} are invoked to derive the  kinetic equation for the evolution of the distribution function of the cooled hadron beam. A qualitative derivation of the MBEC cooling rate using the same approach as the classical stochastic cooling is given in Ref.~\cite{Stupakov:2018HB}.

In this paper, we adopt a framework that differs from what has been used in the literature before. Instead of considering individual hadrons affected by the fields generated by electrons, we look at the dynamics of the fluctuations in both beams. We assume that before the beams start to interact, their density and energy fluctuations can be described as uncorrelated shot noise. In the process of interaction, the fluctuations in the electron and hadron beams establish correlations, and when the beams are recombined in the kicker the fluctuating electric field in the electron beam acts in a way that  decreases the energy spread in the hadron beam. We believe that the language of fluctuations is more appropriate  for the description of the coherent cooling because the interaction involves many particles, in contrast to hadron-electron binary collisions in the incoherent electron cooling.

The paper is organized as follows. In Section~\ref{sec:2} we formulate equations for 1D shot noise when the beam is treated as an ideal gas of non-interacting particles. In Section~\ref{sec:3} we consider the interaction of the hadron and electron beams in the modulator. This interaction is described in general terms of the effective wakefield or, equivalently, impedance. To simplify calculations and to clarify the physical mechanism of the cooling, in Section~\ref{sec:4}, we assume a small dispersion strength of the hadron chicane. In Section~\ref{sec:5} we drop this assumption and derive general expressions for the cooling rate and the energy diffusion in the process of coherent cooling. In Section~\ref{sec:6} we calculate the effective interaction impedance due to the Coulomb interaction of hadrons and electrons in the modulator. In Section~\ref{sec:7} we optimize the strengths of the hadron and electron chicanes and derive the final formula, Eq.~\eqref{eq:67}, for the cooling rate. In Section~\ref{sec:8} we estimate the energy diffusion associated with the cooling and in Section~\ref{sec:9} we apply our formulas to the parameters of the eRHIC collider. In Section~\ref{sec:10} we compare our computer simulations with the theory and we conclude with a discussion of our results in Section~\ref{sec:11}.

We use the Gaussian system of units throughout this paper.

%
\section{Shot noise in beams}\label{sec:2}
%

We consider fluctuations in a beam on the scale that is much smaller than the beam length. Locally, the beam can be treated as having an average distribution function that does not depend on the longitudinal coordinate $z$. Throughout this paper we use the notation $z$ for the longitudinal coordinate inside the bunch, $z=s-v_0t$, where $s$ the longitudinal coordinate in the lab frame and $v_0$ is the nominal beam velocity. The analysis in this section is applicable to both electron and hadron beams, so we do not use indices that indicate a species. In subsequent sections we will use $e$ and $h$ for electron and hadron quantities, respectively.

We denote by $\eta$ the relative energy deviation of a particle in the beam, $\eta=\Delta E/E_0$, where $E_0 = \gamma mc^2$ is the nominal energy. The initial 1D distribution function, before the beam enters the modulator, is
    \begin{align}\label{eq:1}
    f_0(z,\eta) = n_0 F(\eta)+\delta f(z,\eta)
    ,
    \end{align}
where $F(\eta)$ is the averaged energy distribution function normalized by $\int d\eta F(\eta) = 1$, and $n_0$ is the averaged 1D density of the beam (the number of particles per unit length). In this local analysis of fluctuations, the beam is considered as infinitely long, so $F$ does not depend on the coordinate $z$. The function $\delta f(z,\eta)$ describes statistical fluctuations in the beam; it has an average value equal to zero, $\langle\delta f(z,\eta)\rangle=0$. Generally speaking, fluctuations evolve with time or, equivalently, along the beam path $s$, but for brevity we omit the variable $s$ from the arguments of $\delta f$. In what follows, we will only need to calculate $\delta f(z,p)$ at several specific locations along the beam line.  

We define the Fourier transformation of $\delta f$ by the following equations: 
    \begin{align}\label{eq:2}
    \delta \hat f_{k}(\eta)
    &=
    \intinf  dz
    e^{-ikz}
    \delta f(z,\eta)
    ,\qquad
    \delta f(z,\eta) 
    =
    \frac{1}{2\pi}
    \intinf  dk
    e^{ikz}
    \delta \hat f_{k}(\eta)
    .
    \end{align}
If we neglect the electromagnetic interaction between the particles and treat the beam as an ideal gas, according to the kinetic theory of gases~\cite{landau_lifshitz_pk}, the correlator of two functions $\delta f$ taken at different points in the phase space is given by the following formula
    \begin{align}\label{eq:3}
    \langle
    \delta f(z,\eta)
    \delta f(z',\eta')    
    \rangle
    =
    n_0 F(\eta)
    \delta(z-z')
    \delta(\eta-\eta')
    ,
    \end{align}
which, after the Fourier transformation, gives
    \begin{align}\label{eq:4}
    \langle
    \delta\hat f_k(\eta)
    \delta\hat f_{k'}(\eta')    
    \rangle
    =
    2\pi 
    n_0 F(\eta)
    \delta(k+k')
    \delta(\eta-\eta')
    .
    \end{align}
Eqs.~\eqref{eq:3} and~\eqref{eq:4} are the mathematical expressions of the so called \emph{shot noise} in the beam. 
 
Introducing the density fluctuation $\delta n(z)$ as
	\begin{align}\label{eq:5}
	\delta n(z)
	=
	\intinf d\eta\,
	\delta f(z,\eta)
	,
	\end{align}
we find by integrating Eq.~\eqref{eq:3} over $\eta$ and $\eta'$,
	\begin{align}\label{eq:6}
	\langle \delta n(z) \delta n(z') \rangle = n_0\delta(z-z')
	,
	\end{align}
which means that the density fluctuations in the shot noise are uncorrelated. We can also calculate the Fourier spectrum of $\delta n(z)$,
    \begin{align}\label{eq:7}
    \delta \hat n_k
    =
    \int_{-\infty}^\infty
    dz
    e^{-ikz}
    \delta n(z)
    =
    \int_{-\infty}^\infty
    d\eta\,
    \delta\hat f_k(\eta)
    .
    \end{align}
Integrating Eq.~\eqref{eq:4} over $\eta$ and $\eta'$ we obtain
	\begin{align}\label{eq:8}
    \langle \delta \hat n_k \delta \hat n_{k'}\rangle
    =
    2\pi 
    n_0 
    \delta(k+k')
    .
	\end{align}
If we integrate Eq.~\eqref{eq:3} over $\eta'$ and make the Fourier transformation over $z'$ we obtain an expression which will need later:
    \begin{align}\label{eq:9}
    \langle
    \delta f(z,\eta)
    \delta \hat n_k    
    \rangle
    =
    n_0 F(\eta)
    e^{-ikz}
    .
    \end{align}

%
\section{Dynamics of fluctuations in the hadron beam}\label{sec:3}
%

We now consider the dynamics of fluctuations in the hadron beam as it propagates through the cooling section. We assume that the initial distribution function for hadrons is given by Eq.~\eqref{eq:1} and the beam is in a state with uncorrelated shot noise as described in the previous Section. To distinguish the initial fluctuational part of the distribution function from its final counterpart we will change the notation $\delta f$ in Eq.~\eqref{eq:1} to $\delta f^{(M)}$ (M for the modulator). The beam first interacts with electrons in the modulator where each hadron creates a perturbation in the electron beam. This perturbation is localized in a small vicinity of the hadron. Strictly speaking, electrons also perturb the hadron beam in this interaction, but for now we neglect this effect in our analysis. After passing through the modulator, the hadron beam goes through a chicane with the dispersion characterized by the $R_{56}^{(h)}$ element of the transport matrix, for which we will use a simplified notation $R_h$. Passage through the chicane $R_h$ introduces a phase-space transformation $(z,\eta)\to(z',\eta')$: $z' = z + R_h\eta$, $\eta'=\eta$, and changes the initial hadron distribution function $f_0$ in the modulator into a different function, $f_1$, in the kicker, $f_0 \to f_1$. The new distribution function is obtained by expressing the old arguments through the new ones\footnote{Here we implicitly assume that the ion motion from the kicker to the modulator is Hamiltonian and the distribution function remains constant along the trajectories in the phase space.}:
    \begin{align}\label{eq:10}
    f_1(z,\eta)
    = 
    f_0(z-R_h\eta,\eta)
    =
    n_{0h} F_h(\eta)
    +
    \delta f^{(M)}(z-R_h\eta,\eta)
    ,
    \end{align}
where $F_h$ is the averaged energy distribution function of hadrons and $n_{0h}$ is the linear density of particles in the hadron beam. The hadron beam then goes into the kicker where it interacts with the electron beam again. This interaction changes the relative energy of the hadrons located at coordinate $z$ by $\Delta \eta^{(h)}(z)$, $\eta'=\eta+\Delta \eta^{(h)}(z)$ (we will discuss the specific form of $\Delta \eta^{(h)}(z)$ below). This results in a new hadron distribution function after the kicker, $f_1 \to f_2$, 
    \begin{align}\label{eq:11}
    f_2(z,\eta)
    &=
    f_1(z,\eta-\Delta \eta^{(h)})
     \nonumber\\
    &   
    =
    n_{0h} F_h(\eta-\Delta \eta^{(h)})
    +
    \delta f^{(M)}(z-R_h\eta
    +
    R_h\Delta \eta^{(h)},\eta
    -
    \Delta \eta^{(h)})
    .
    \end{align}

We will now look more closely at the hadron-electron interaction \emph{in the kicker}. We assume that this interaction can be characterized by an effective wakefield $w(z)$ such that the energy change $\Delta E(z)$  of a hadron located at point $z$ in the beam after a passage through the kicker is
    \begin{align}\label{eq:12}
    \Delta E(z)
    =
    (Ze)^{2}
    \int_{-\infty}^{\infty}
    w(z-z')
    \delta n^{(M)}(z')
    dz'
    ,
    \end{align}
(we use the convention that the positive wake corresponds to the energy gain), where $\delta n^{(M)}$ is the hadron density fluctuation \emph{in the modulator}, $\delta n^{(M)}(z) = \int \delta f^{(M)}(z,\eta)d\eta$, and $Ze$ is the hadron charge (throughout this paper we use the notation $e$ for the \emph{positive} elementary charge). It is also convenient to introduce the impedance ${\cal Z}(k)$ related to the wake through the equation
    \begin{align}\label{eq:13}
    {\cal Z}(k)
    =
    -
    \frac{1}{c}
    \int_{-\infty}^{\infty}
    dz\,
    w(z)
    e^{-ikz}
    ,\qquad
    w(z)
    =
    -
    \frac{c}{2\pi}
    \int_{-\infty}^{\infty}
    dk\,
    {\cal Z}(k)
    e^{ikz}
    .
    \end{align}
Being a Fourier transform of a real function, the real and imaginary parts of ${\cal Z}(k)$ are respectively even and odd functions of $k$, ${\cal Z}(-k)={\cal Z}^*(k)$. The relative energy change of a hadron at coordinate $z$ can now be written as 
    \begin{align}\label{eq:14}
    \Delta \eta^{(h)}(z)
    =
    -
    \frac{r_{h}c}{2\pi \gamma}
    \int_{-\infty}^\infty
    {dk}
    {\cal Z}(k)
    \delta \hat n_{k}^{(M)}
    e^{ikz}
    ,
    \end{align}
where $\Delta \eta^{(h)} = \Delta E/\gamma m_h c^2$, $\gamma m_h c^2$ is the nominal energy of the beam, $\delta\hat n_{k}^{(M)}$  is the Fourier transform of $\delta n_h^{(M)}(z)$, and $r_h = (Ze)^2/m_hc^2$.  Introducing the Fourier transform $\Delta \hat \eta_k^{(h)}$ as defined by Eqs.~\eqref{eq:2}, we obtain
    \begin{align}\label{eq:15}
    \Delta \hat \eta_k^{(h)}
    =
    -
    \frac{r_hc}{\gamma}
    {\cal Z}(k)
    \delta \hat n_{k}^{(M)}
    .
    \end{align}
It is important to remember that here $\delta \hat n_{k}^{(M)}$ is associated with the density fluctuations {in the modulator}---the place where the hadron fluctuations are imprinted on the electrons. These fluctuation should be calculated with the initial fluctuational part of the distribution function $\delta f^{(M)}(z,\eta)$. When hadrons arrive to the kicker having been longitudinally displaced in the chicane their distribution function changes to $f_2$ given by Eq.~\eqref{eq:11}. The hadron fluctuations in the kicker differ from the initial noise in the modulator described by equations in Section~\ref{sec:2}.

%
\section{Coherent cooling in the limit of a small value of the chicane strength}\label{sec:4}
%

As a result of the passing through the cooling section, the distribution function of the hadron beam changes. Introducing the difference 
    \begin{align}\label{eq:16}
    \Delta f(z,\eta) 
    =  
    f_2(z,\eta)-n_{0h} F_h(\eta),
    \end{align}
where $f_2(z,\eta)$ is the distribution function after the kicker, we should not expect that the averaged value $\langle\Delta f\rangle$ vanishes, in contrast to the zero value of $\langle\delta f^{(M)}\rangle$ in the initial state. We associate the average value of $\Delta f$ with the change of the averaged distribution function in one revolution in the ring:
    \begin{align}\label{eq:17}
    n_{0h}T
    \frac{\p F_h}{\p t}
    =
    \langle\Delta f\rangle
    ,
    \end{align}
where $T$ is the revolution period. As we will see below, this equations describes a gradual decrease of the hadron beam energy spread due to the coherent cooling.

We will now calculate $\langle\Delta f\rangle$. To simplify analysis, in this Section we will assume that $R_h$ is small and  use the Taylor expansion in Eq.~\eqref{eq:11} keeping terms linear in $R_h$. In addition, we will use the smallness of the fluctuations in the beam and treat $\Delta \eta^{(h)}$ and $\delta f^{(M)}$ as small quantities $\sim\epsilon$ and neglecting terms of order $\epsilon^3$ and higher. Using Eq.~\eqref{eq:11}, we find
    \begin{align}\label{eq:18}
    \langle\Delta f\rangle
    &\approx
    \langle
    \frac{1}{2}
    n_{0h}
    (\Delta \eta^{(h)})^{2}
    F_h''(\eta)
    -
    \Delta \eta^{(h)}
    \p_\eta
    \delta f^{(M)}
    +
    R_h\eta\Delta \eta^{(h)}
    \p_{z\eta}
    \delta f^{(M)}
    +
    R_h\Delta \eta^{(h)}
    \p_{z}
    \delta f^{(M)}
    \rangle
    \nonumber\\
    &=
    \frac{1}{2}
    n_{0h}
    \langle
    (\Delta \eta^{(h)})^{2}
    \rangle
    F_h''(\eta)
    -
    \langle
    \Delta \eta^{(h)}
    \p_\eta
    \delta f^{(M)}
    \rangle
    +
    R_h
    \langle
    \Delta \eta^{(h)}
    \p_{z\eta}
    (
    \eta\delta f^{(M)}
    )
    \rangle
    ,
    \end{align}
where we have omitted the arguments $(z,\eta)$ in the function $\delta f^{(M)}$ and took into account that $\langle \delta f^{(M)}\rangle = 0$ and $\langle \Delta \eta^{(h)} \rangle =0$. We expect that the last term on the right-hand side of this equation is associated with the cooling because it is proportional to the product of the chicane strength and the energy change $\Delta \eta^{(h)}$  in the interaction. With the help of Eqs.~\eqref{eq:14} and~\eqref{eq:9} we can write this term as
    \begin{align}\label{eq:19}
    R_h
    \langle
    \Delta \eta^{(h)}
    \p_{z\eta}
    (\eta\delta f^{(M)})
    \rangle
    &=
    -
    \frac{R_hr_hc}{2\pi \gamma}
    \int_{-\infty}^\infty
    {dk}
    {\cal Z}(k)
    e^{ikz}
    \p_{z\eta}
    (
    \eta
    \langle
    \delta \hat n_k^{(M)}
    \delta f
    \rangle
    )
    \nonumber\\&
    =
    \frac{R_hr_hc}{2\pi \gamma}
    n_{0h} 
    \p_\eta(\eta F_h)
    \int_{-\infty}^\infty
    ik{\cal Z}(k)dk
    .
    \end{align}
From Eq.~\eqref{eq:13} it follows that
    \begin{align}\label{eq:20}
    \int_{-\infty}^{\infty}
    ik{\cal Z}(k)dk
    =   
    -
    \frac{2\pi}{c}
    w'_{l}(0)
    ,
    \end{align}
so that the right-hand side of Eq.~\eqref{eq:19} can be written as
    \begin{align}\label{eq:21}
    -
    w'_{l}(0)
    \frac{R_hr_h}{\gamma}
    n_{0h} \p_\eta(\eta F_h)
    .
    \end{align}
Substituting this into the right-hand side of Eq.~\eqref{eq:17} yields
    \begin{align}\label{eq:22}
    \frac{\p F_h}{\p t}
    =
    -
    w'_{l}(0)
    \frac{R_hr_h}{T\gamma}
    \frac{\p (\eta F_h)}{\p \eta}
    .
    \end{align}
To see how Eq.~\eqref{eq:22} describes evolution of the rms energy spread of the beam, $\sigma_h$,  with time we multiply it by $\eta^2$ and integrate over $\eta$,
    \begin{align}\label{eq:23}
    \frac{d \sigma_h^2}{d t}
    =
    \intinf
    d\eta\,
    \eta^2
    \frac{\p F_h}{\p t}
    =
    -
    w'_{l}(0)
    \frac{R_hr_h}{T\gamma}
    \int
    d\eta\,
    \eta^2
    \frac{\p \eta F_h}{\p \eta}
    =
    2
    w'_{l}(0)
    \frac{R_hr_h}{T\gamma}
    \sigma_h^2
    .
    \end{align}
Hence, for $w'_{l}(0)R_h<0$, we have an exponential cooling with the time constant
    \begin{align}\label{eq:24}
    t_\mathrm{c}^{-1}
    \equiv
    \sigma_h^{-2}
    \frac{d \sigma_h^2}{d t}
    =
    2
    \frac{r_h}{T\gamma}
    |R_hw'_{l}(0)|
    .
    \end{align}
The physical mechanism of this effect is the following~\cite{Ratner2013}. Assuming for simplicity  $w_l(0)=0$ and also $R_h>0$ and $w'_{l}(0)<0$, we see that hadrons passing through the chicane with the energy higher than the nominal one are shifted in the forward direction by positive $R_h$, and their energy is decreased by the negative wake in the kicker. Particles with the energy smaller then the nominal are shifted by the chicane backward, and their energy is increased by the positive wake. As a result, the repetitive passages through the cooling system lead to a gradual decrease of the energy spread of the hadron beam.

We now consider the second term on the right-hand side of Eq.~\eqref{eq:18}. Again using Eqs.~\eqref{eq:14} and~\eqref{eq:9} we find,
    \begin{align}\label{eq:25}
    -
    \p_\eta
    \langle
    \Delta \eta^{(h)}
    \delta f^{(M)}
    \rangle
    &=
    \frac{r_hc}{2\pi \gamma}
    \int_{-\infty}^\infty
    {dk}
    {\cal Z}(k)
    e^{ikz}
    \p_\eta
    \langle
    \delta \hat n_{k}^{(M)}
    \delta f^{(M)}
    \rangle
    \nonumber\\&
    =
    n_{0h}
    F_h'(\eta)
    \frac{r_hc}{2\pi \gamma}
    \intinf
    {dk}
    {\cal Z}(k)
    .
    \end{align}
The integral $\intinf {dk}    {\cal Z}(k)$ can be expressed through the value of the wake at the origin, $w(0)$, and Eq.~\eqref{eq:25} can be written as
    \begin{align}\label{eq:26}
    -
    n_{0h}F_h'(\eta)
    \frac{r_h}{\gamma}
    w(0)
    .
    \end{align}
This term can be interpreted as a change of the particle energy due to its own wake, if the wake at the origin is not zero, $w(0)\ne 0$. When the process is repeated every revolution period $T$, this term contributes to the time evolution equation for $F_h$:
    \begin{align}\label{eq:27}
    \frac{\p F_h}{\p t}
    =
    -
    \frac{r_h}{T\gamma}
    w(0)
    \frac{\p F_h}{\p \eta}
    .
    \end{align}
Multiplying this equation by $\eta$ and integrating it over $\eta$ gives the following equation for the average rate of the energy loss\footnote{In the case when the wake $w(z)$ is due to the interaction with accelerating cavities, this energy is lost to the excitation of cavity modes---the well known effect of the beam loading.}    
    \begin{align}\label{eq:28}
    \frac{1}{E_0}
    \frac{d E_0}{d t}
    =    
    \frac{r_h}{T\gamma}    
    w(0)
    .
    \end{align}
Clearly this term is of no interest for the cooling process. Moreover, for the space charge interaction considered in Section~\ref{sec:6} the value of the wake at the origin is equal to zero, and this term vanishes.

Finally, the first term in Eq.~\eqref{eq:18} can be written as follows:
    \begin{align}\label{eq:29}
    \frac{1}{2}
    n_{0h}
    F_h''(\eta)
    \langle
    (\Delta \eta^{(h)})^{2}
    \rangle
    &=
    \frac{1}{2}
    n_{0h}
    F_h''(\eta)
    \left(
    \frac{r_hc}{2\pi \gamma}
    \right)^2
    \int_{-\infty}^\infty
    {dk}
    {dk'}
    {\cal Z}(k)
    {\cal Z}(k')
    e^{ikz+ik'z}
    \langle
    \delta \hat n_{k'}^{(M)}
    \delta \hat n_k^{(M)}
    \rangle
    \nonumber\\
    &=
    \frac{1}{4\pi}
    n_{0h}^2
    F_h''(\eta)
    \left(
    \frac{r_hc}{ \gamma}
    \right)^2
    \int_{-\infty}^\infty
    {dk}
    |{\cal Z}(k)|^2
    ,
    \end{align}
where we have used the relation~\eqref{eq:8}. Its contribution to the time derivative of $F_h$ is a diffusion-like term:
    \begin{align}\label{eq:30}
    \frac{\p F_h}{\p t}
    =
    D
    \frac{\p^2 F_h}{\p^2 \eta}
    ,
    \end{align}    
with the diffusion coefficient
    \begin{align}\label{eq:31}
    D
    =
    \frac{n_{0h}}{4\pi T}
    \left(
    \frac{r_hc}{ \gamma}
    \right)^2
    \int_{-\infty}^\infty
    {dk}
    |{\cal Z}(k)|^2
    .
    \end{align}
This diffusion is caused by the shot noise in the hadron beam that is transferred through the interaction with the electron beam and then applied back to the hadrons in the kicker. 

The averaged energy loss, Eq.~\eqref{eq:28}, and the diffusion coefficient, Eq.~\eqref{eq:31}, can also be derived in a single-particle treatment of the beam as shown in Appendix~\ref{sec:app1}.

Taking into account both the cooling and diffusion, we need to combine Eqs.~\eqref{eq:22} and~\eqref{eq:30},
    \begin{align}\label{eq:32}
    \frac{\p F_h}{\p t}
    =
    \frac{1}{2t_\mathrm{c}}
    \frac{\p (\eta F_h)}{\p \eta}
    +
    D
    \frac{\p^2 F_h}{\p^2 \eta}
    .
    \end{align}
Multiplying this equation by $\eta^2$ and integrating it over $\eta$, as was done in Eq.~\eqref{eq:23}, we obtain
    \begin{align}\label{eq:33}
    \frac{d\sigma_h^2}{dt}
    =
    -
    \frac{\sigma_h^2}{t_\mathrm{c}}
    +
    2D
    .
    \end{align}
From this equation it follows that for a cooling effect to prevail over the diffusion, the value of $D$ should not be too large,
    \begin{align}\label{eq:34}
    D
    <
    \frac{\sigma_h^2}{2t_\mathrm{c}}
    .
    \end{align}
In the opposite limit, the heating due to the diffusion overcomes the cooling and the initial energy spread of the hadron beam grows with time, ${d\sigma_h^2}/{dt}>0$.

It follows from Eq.~\eqref{eq:24} that for a faster cooling one would like to have a larger value of the dispersion strength $R_h$. However, our analysis in this section assumed a small $R_h$ and hence cannot be applied to arbitrary values of $R_h$. The assumption of the small chicane strength will be dropped in the next section.

%
\section{Cooling for arbitrary value of the chicane strength}\label{sec:5}
%

We now return to Eq.~\eqref{eq:11} and repeat the derivation of $\Delta f$ without making an assumption that $R_h\eta$ is small (but still assuming the smallness of $R_h\Delta \eta^{(h)}$ which is proportional to the small quantity $\Delta \eta^{(h)}$). Instead of Eq.~\eqref{eq:18} we now obtain
    \begin{align}\label{eq:35}
    \langle\Delta f\rangle
    &=
    \frac{1}{2}
    n_{0h}
    \langle
    (\Delta \eta^{(h)})^{2}
    \rangle
    F_h''(\eta)
    -
    \langle
    \Delta \eta^{(h)}
    \p_2\delta f^{(M)}(z-R_h\eta,\eta)
    \rangle
    +
    \langle
    R_h\Delta \eta^{(h)}
    \p_z\delta f^{(M)}(z-R_h\eta,\eta)
    \rangle
    ,
    \end{align}
where $\p_2\delta f^{(M)}$ denotes the partial derivative with respect to the second arguments of $\delta f^{(M)}$. The first term here, as the first term in Eq.~\eqref{eq:18}, is responsible for the energy diffusion~\eqref{eq:30}. This term will not be considered below. The cooling effect is due to the second and the last terms that involve $R_h$, but they also include the average energy loss of the hadrons if $w_l(0)\ne 0$, as described by Eq.~\eqref{eq:27}. 

We begin with the calculation of the average of the last term in Eq.~\eqref{eq:35}:
    \begin{align}\label{eq:36}
    R_h
    \langle
    \Delta \eta^{(h)}
    \p_z\delta f^{(M)}(z-R_h\eta,\eta)
    \rangle
    &=
    -
    \frac{R_hr_hc}{2\pi \gamma}
    \int_{-\infty}^\infty
    {dk}
    {\cal Z}(k)
    e^{ikz}
    \p_{z}
    \langle
    \delta \hat n_{k}^{(M)}
    \delta f^{(M)}(z-R_h\eta,\eta)
    \rangle
    .
    \end{align}
Similar to the derivation of Eq.~\eqref{eq:9} it is easy to find that
    \begin{align}\label{eq:37}
    \langle
    \delta \hat n_{k}^{(M)}
    \delta f^{(M)}(z-R_h\eta,\eta)
    \rangle
    =
    n_{0h} F_h(\eta)
    e^{-ik(z-R_h\eta)}
    ,
    \end{align}
which gives
    \begin{align}\label{eq:38}
    R_h
    \langle
    \Delta \eta^{(h)}
    \p_z\delta f^{(M)}(z-R_h\eta,\eta)
    \rangle
    &=
    in_{0h} F_h(\eta)
    \frac{R_hr_hc}{2\pi \gamma}
    \int_{-\infty}^\infty
    {dk}
    k{\cal Z}(k)
    e^{ikR_h\eta}
    .
    \end{align}
For the second term in Eq.~\eqref{eq:35} we have
    \begin{align}\label{eq:39}
    -
    \langle
    \Delta \eta^{(h)}
    \p_2\delta f^{(M)}(z-R_h\eta,\eta)
    \rangle
    &=
    \frac{r_hc}{2\pi \gamma}
    \int_{-\infty}^\infty
    {dk}
    {\cal Z}(k)
    e^{ikz}
    \langle
    \delta \hat n_{k}^{(M)}
    \p_2\delta f^{(M)}(z-R_h\eta,\eta)
    \rangle
    .
    \end{align}
Again, following the derivation of Eq.~\eqref{eq:9}, we find
    \begin{align}\label{eq:40}
    \langle
    \delta \hat n_k^{(M)}
    \p_2\delta f^{(M)}(z-R_h\eta,\eta)
    \rangle
    =
    n_{0h} F_h'(\eta)
    e^{-ik(z-R_h\eta)}
    ,
    \end{align}
which gives
    \begin{align}\label{eq:41}
    -
    \langle
    \Delta \eta^{(h)}
    \p_2\delta f^{(M)}(z-R_h\eta,\eta)
    \rangle
    &=
    \frac{r_hc}{2\pi \gamma}
    n_{0h} F_h'(\eta)
    \int_{-\infty}^\infty
    {dk}
    {\cal Z}(k)
    e^{ikR_h\eta}
    .
    \end{align}
Adding the right-hand sides of Eqs.~\eqref{eq:38} and~\eqref{eq:41}, we will subtract the effect of the wake at the origin, Eq.~\eqref{eq:25}, to obtain
    \begin{align}\label{eq:42}
    n_{0h}
    \frac{r_hc}{2\pi \gamma}
    \int_{-\infty}^\infty
    {dk}
    {\cal Z}(k)
    \left[
    ikR_hF_h(\eta)e^{ikR_h\eta}
    +
    F_h'(\eta)(e^{ikR_h\eta}-1)
    \right]
    .
    \end{align}
In the limit of small $R_h$, this expression reduces to Eq.~\eqref{eq:19} and hence it generalizes the cooling term on the right-hand side of the kinetic equation~\eqref{eq:22} to arbitrary values of $R_h$. 

Note that in this regime the right-hand side of the kinetic equation for function $F_h$ differs from a simple form, Eq.~\eqref{eq:32}, valid in the limit of small $R_h$. In particular, the cooling term in this equation is not equal any more to the derivative $\p(\eta F_h)/\p\eta$ divided by the twice the cooling time --- it will now involve a more complicated expression with the integral from Eq.~\eqref{eq:42}. However, we still can define the cooling time $t_\mathrm{c}$ as an inverse rate of change of $\sigma_h^2$,
    \begin{align*}
    t_\mathrm{c}^{-1}
    =
    \left(
    \intinf
    d\eta\,
    \eta^2
    \frac{\p F_h}{\p t}
    \right)
    \left(
    \intinf
    d\eta\,
    \eta^2
    F_h
    \right)^{-1}
    .
    \end{align*}
Repeating the derivation of Eq.~\eqref{eq:23} we find,
    \begin{align}\label{eq:43}
    t_\mathrm{c}^{-1}
    =
    -
    \frac{r_hc}{2\pi \gamma\sigma_h^2T}
    \int_{-\infty}^\infty
    {dk}
    {\cal Z}(k)
    \int_{-\infty}^\infty
    \eta^2d\eta
    \left[
    ikR_hF_h(\eta)e^{ikR_h\eta}
    +
    F_h'(\eta)(e^{ikR_h\eta}-1)
    \right]
    ,
    \end{align}
assuming that the right-hand side of this equation is positive. The integrand in this expression as a function of $k$ has the same symmetry as ${\cal Z}(k)$---changing the sign of $k$  makes it a complex conjugate. Using this symmetry we can re-write Eq.~\eqref{eq:43} in explicitly real form,
    \begin{align}\label{eq:44}
    t_\mathrm{c}^{-1}
    =
    -
    \frac{r_hc}{\pi \gamma\sigma_h^2T}
    \Re
    \int_{0}^\infty
    {dk}
    {\cal Z}(k)
    \int_{-\infty}^\infty
    \eta^2d\eta
    \left[
    ikR_hF_h(\eta)e^{ikR_h\eta}
    +
    F_h'(\eta)(e^{ikR_h\eta}-1)
    \right]
    .
    \end{align}

To proceed further, we need to specify the impedance ${\cal Z}(k)$ and then to find $R_h$ that minimizes the cooling time. 

%
\section{Impedance for MBEC}\label{sec:6}
%

We will now discuss the effective impedance $\cal Z$ of the MBEC cooling method in a simple setup shown in Fig.~\ref{fig:1}. This impedance is generated when the electron beam first interacts with hadrons in the modulator, travels through its chicane, $R_{56}^{(e)}$, and then interacts with hadrons again in the kicker section. For the hadron-electron interaction we will adopt a model in which the interaction is treated as if a hadron were a disk of charge $Ze$ with an axisymmetric Gaussian radial distribution with the rms transverse size equal to the rms transverse size of the beam. The electron is also modeled by a Gaussian disk of charge $-e$ with the same transverse profile. We believe that this model is more accurate that the one developed in Ref.~\cite{Ratner2013} where the interaction was treated as between a uniformly charged disk and a point charge on the axis of the beam. A similar Gaussian-to-Gaussian interaction model was used in 1D simulations of a longitudinal space charge amplifier in Ref.~\cite{dohlus2011sy}. 

\begin{figure}[htb]
\centering
\includegraphics[width=0.6\textwidth, trim=0mm 0mm 0mm 0mm, clip]{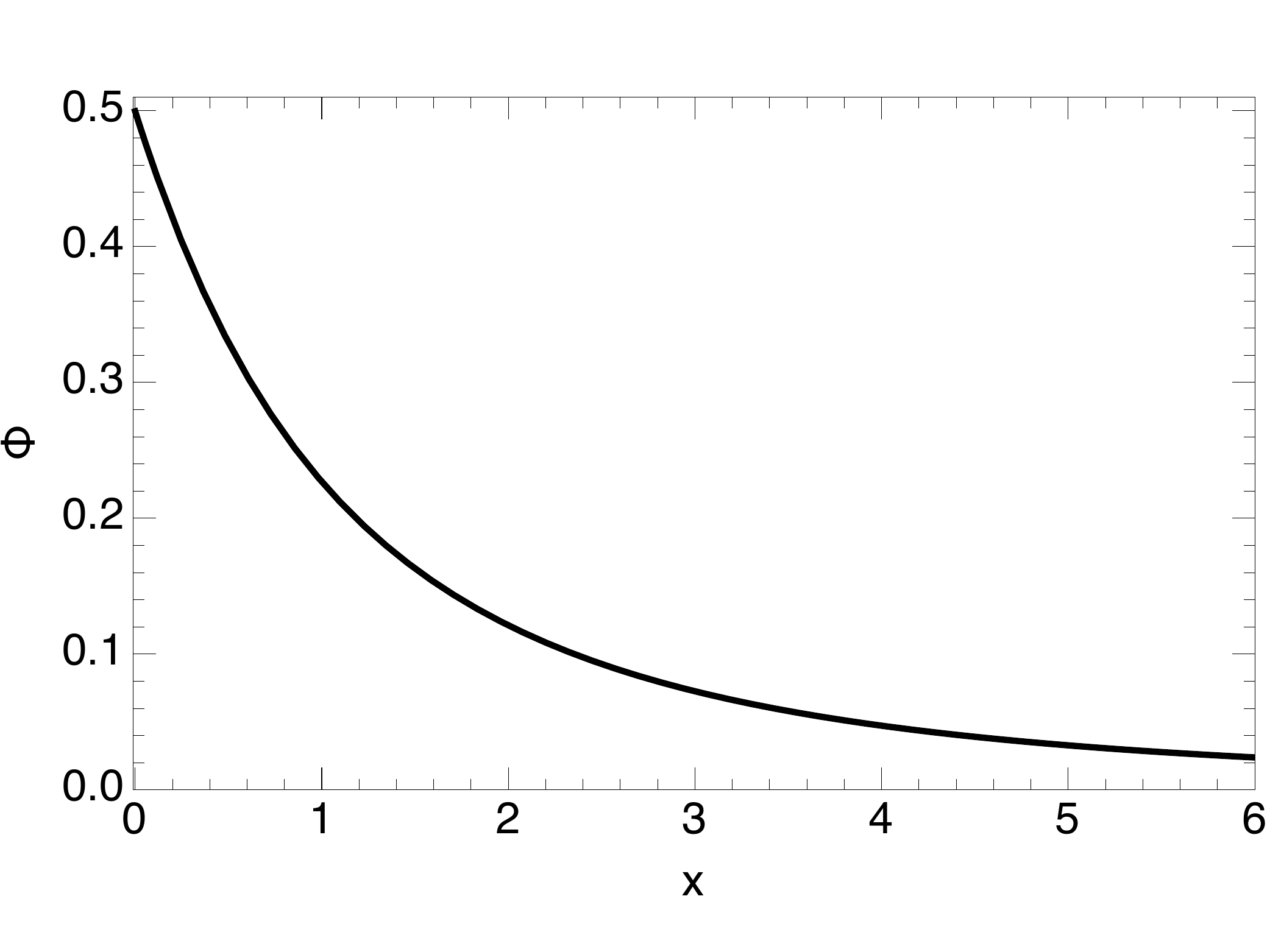}
\caption{Function $\Phi(x)$ for positive values of the argument. }
\label{fig:2}
\end{figure}

In this model, a hadron of charge $Ze$ at the origin of the coordinate system exerts a force $f_z$  on an electron at coordinate $z$,
    \begin{align}\label{eq:45}
    f_z(z)
    =
    -
    \frac{Ze^2}{\Sigma^2}
    \Phi
    \left(\frac{z\gamma}{\Sigma}\right)
    ,
    \end{align}
where $\Sigma$ is the rms beam radius and the function $\Phi$ is defined by the following expression~\cite{geloni07ssy},
    \begin{align}\label{eq:46}
    \Phi(x)
    =
    \frac{1}{2}
    \left[
    \frac{x}{|x|}
    -
    \frac{x\sqrt{\pi}}{2}
    \exp
    \left(
    \frac{1}{4}x^2
    \right)
    \mathrm{erfc}
    \left(
    \frac{1}{2}|x|
    \right)
    \right]
    ,
    \end{align}
with erfc the complementary error function.  The function $\Phi$ is odd, $\Phi(-x) = -\Phi(x)$; its plot for positive $x$ is shown in Fig.~\ref{fig:2}.
Neglecting the relative longitudinal displacements of a hadron and an electron in the modulator, the force~\eqref{eq:45} causes the relative energy change $G_\eta$ in an electron located at coordinate $z$,
    \begin{align}\label{eq:47}
    G_\eta(z)
    =
    -
    \frac{Zr_eL_m}{\gamma \Sigma^2}
    \Phi
    \left(\frac{z\gamma}{\Sigma}\right)
    ,
    \end{align}
where $L_m$ is the length of the modulator and $r_e=e^2/m_ec^2$ is the classical electron radius. 

The function $G_\eta$ can also be considered as a Green function for the energy modulation of electrons induced by a delta-function density perturbation in the hadron beam. With the help of this Green function an energy modulation $\Delta\eta^{(e)}$ in the electron beam in the modulator induced by a 1D density modulation if the hadron beam $\delta n^{(M)}(z)$ can be written as 
    \begin{align}\label{eq:48}
    \Delta\eta^{(e)}(z)
    =
    \int_{-\infty}^\infty
    dz'
    \delta n^{(M)}(z')
    G_\eta(z-z')
    .
    \end{align}

We denote the averaged electron distribution function by $n_{e0}F_e(\eta)$. In what follows, we neglect fluctuations in the electron beam and hence do not introduce the $\delta f$ term, as in Eq.~\eqref{eq:1}, for electrons. After the interaction with the hadrons, the electron energy distribution becomes $n_{e0}F_e(\eta-\Delta \eta^{(e)} (z))$. The chicane in the electron path with the $R_{56}^{(e)}\equiv R_e$ matrix element shifts electrons in the longitudinal direction, $z'=z+R_e\eta$, and hence the distribution function after the electron chicane becomes
    \begin{align}\label{eq:49}
    n_{0e}F_e[\eta-\Delta\eta^{(e)}(z-R_e\eta)]
    .
    \end{align}
Let us now calculate the density perturbation of the electrons after the chicane, $\delta n^{(e)}$, in the kicker:
    \begin{align}\label{eq:50}
    \delta n^{(e)}(z)
    &=
    n_{0e}
    \int_{-\infty}^\infty
    d\eta
    [
    F_e(\eta-\Delta\eta^{(e)}(z-R_e\eta))
    -
    F_e(\eta)
    ]
    \nonumber\\
    &\approx
    -
    n_{0e}
    \int_{-\infty}^\infty
    d\eta
    F'_e(\eta)
    \Delta\eta^{(e)}(z-R_e\eta)
    ,
    \end{align}
where we have used the Taylor expansion keeping only a linear term in $\Delta\eta^{(e)}$. For the Fourier transform of the electron density perturbation we find
    \begin{align}\label{eq:51}
    \delta \hat n_{k}^{(e)}
    &=
    \int_{-\infty}^\infty
    dz
    e^{-ikz}
    \delta n^{(e)}(z)
    =
    -
    n_{0e}
    g(k)
    \Delta\hat\eta_k^{(e)}
    ,
    \end{align}
where
    \begin{align}\label{eq:52}
    g(k)
    =
    \int_{-\infty}^\infty
    d\eta
    F'_e(\eta)
    e^{-ikR_e\eta}
    ,
    \end{align}
and $\Delta\hat\eta_k^{(e)}$ is the Fourier transform of $\Delta\eta^{(e)}(z)$. For a Gaussian distribution function of electrons, $F_e= (\sqrt{2\pi}\sigma_{e})^{-1}e^{-\eta^2/2\sigma_{e}^2}$, with $\sigma_{e}$ is the rms relative energy spread, we have
    \begin{align}\label{eq:53}
    g(k)
    =
    ikR_e
    e^{-k^2R_e^2\sigma_{e}^2/2}
    .
    \end{align}
Using Eqs.~\eqref{eq:47} and~\eqref{eq:48} we find that
    \begin{align}\label{eq:54}
    \Delta\hat\eta_k^{(e)}
    &=
    -
    Z
    \zeta(k)
    \delta\hat{n}_{k}^{(M)}
    ,
    \end{align}
with
    \begin{align}\label{eq:55}
    \zeta(k)
    &\equiv
    -
    \frac{1}{Z}
    \int_{-\infty}^\infty
    dz
    e^{-ikz}
    G_\eta(z)
    =
    -
    \frac{2ir_eL_m}{\gamma^2 \Sigma}
    H
    \left(
    \frac{k\Sigma}{\gamma}
    \right)
    ,
    \end{align}
and
    \begin{align}\label{eq:56}
    H(x)
    =
    \int_{0}^\infty
    d\xi\,
    \Phi(\xi)
    \sin(x\xi)
    .
    \end{align}
Substituting Eq.~\eqref{eq:54} into Eq.~\eqref{eq:51} we obtain
    \begin{align}\label{eq:57}
    \delta \hat n_{k}^{(e)}
    &=
    Zn_{e0}
    g(k)
    \zeta(k)
    \delta\hat{n}_{k}^{(M)}
    .
    \end{align}

Having found the electron density perturbation we can now calculate the longitudinal force acting on hadrons in the kicker (after the chicane). This is the force that changes the hadron energy. For this force, we will use the same model as above replacing an electron by a disk with a Gaussian distribution with the rms size $\Sigma$ and using Eq.~\eqref{eq:45} (we assume the same transverse size of both beams in the kicker as in the modulator). Following the derivation of Eqs.~\eqref{eq:54}-\eqref{eq:57} it is then straightforward to derive the following formula for the Fourier component of the force $\hat{f}_{zk}$ acting on the hadrons,
    \begin{align}\label{eq:58}
    \hat{f}_{zk}
    =
    \frac{2 i Z^2e^2 n_{0e}}{\Sigma \gamma}
    g(k)
    \zeta(k)
    H
    \left(
    \frac{k\Sigma}{\gamma}
    \right)
    \delta\hat{n}_{k}^{(M)}
    .
    \end{align}
Again, neglecting the relative motion of electrons and hadrons in the kicker, we multiply $\hat{f}_{zk}$ by the length of the kicker $L_k$ and divide it by $\gamma m_h c^2$ to obtain the Fourier component of the energy change $\Delta\hat \eta_k^{(h)}$,
    \begin{align}\label{eq:59}
    \Delta\hat \eta_k^{(h)}
    =
    \frac{2 i e^2 Z^2 n_{0e}L_k}{\Sigma \gamma^2m_h c^2}
    g(k)
    \zeta(k)
    H
    \left(
    \frac{k\Sigma}{\gamma}
    \right)
    \delta\hat{n}_{k}^{(M)}
    .
    \end{align}
Comparing this formula with Eq.~\eqref{eq:15} we find the effective impedance ${\cal Z}(k)$ for the MBEC cooling section,
    \begin{align}\label{eq:60}
    {\cal Z}
    &=
    -
    \frac{2 i  n_{0e}L_k}{c\Sigma \gamma}
    g(k)
    \zeta(k)
    H
    \left(
    \frac{k\Sigma}{\gamma}
    \right)
    \nonumber\\
    &=
    -
    \frac{4 i I_eL_mL_k}{c\Sigma^2 \gamma^3I_A\sigma_{e}}
    q_e\varkappa
    e^{-\varkappa^2q_e^2/2}
    H^2
    \left(
    \varkappa
    \right)
    ,
    \end{align}
where we have introduced the dimensionless parameters
    \begin{align}\label{eq:61}
    \varkappa
    =
    \frac{k\Sigma}{\gamma}
    ,\qquad
    q_e
    =
    \frac{R_e\sigma_{e}\gamma}{\Sigma}
    ,
    \end{align}
and used the electron beam current $I_e=en_{0e}c$ and the Alfv\'{e}n current $I_A = m_ec^3/e\approx 17$ kA. This impedance is purely imaginary and the corresponding wake $w(z)$, as was already indicated above, has a zero value at the origin.
    
%
\section{Maximization of the cooling rate}\label{sec:7}
%

\begin{figure}[htb]
\centering
\includegraphics[width=0.6\textwidth, trim=0mm 0mm 0mm 0mm, clip]{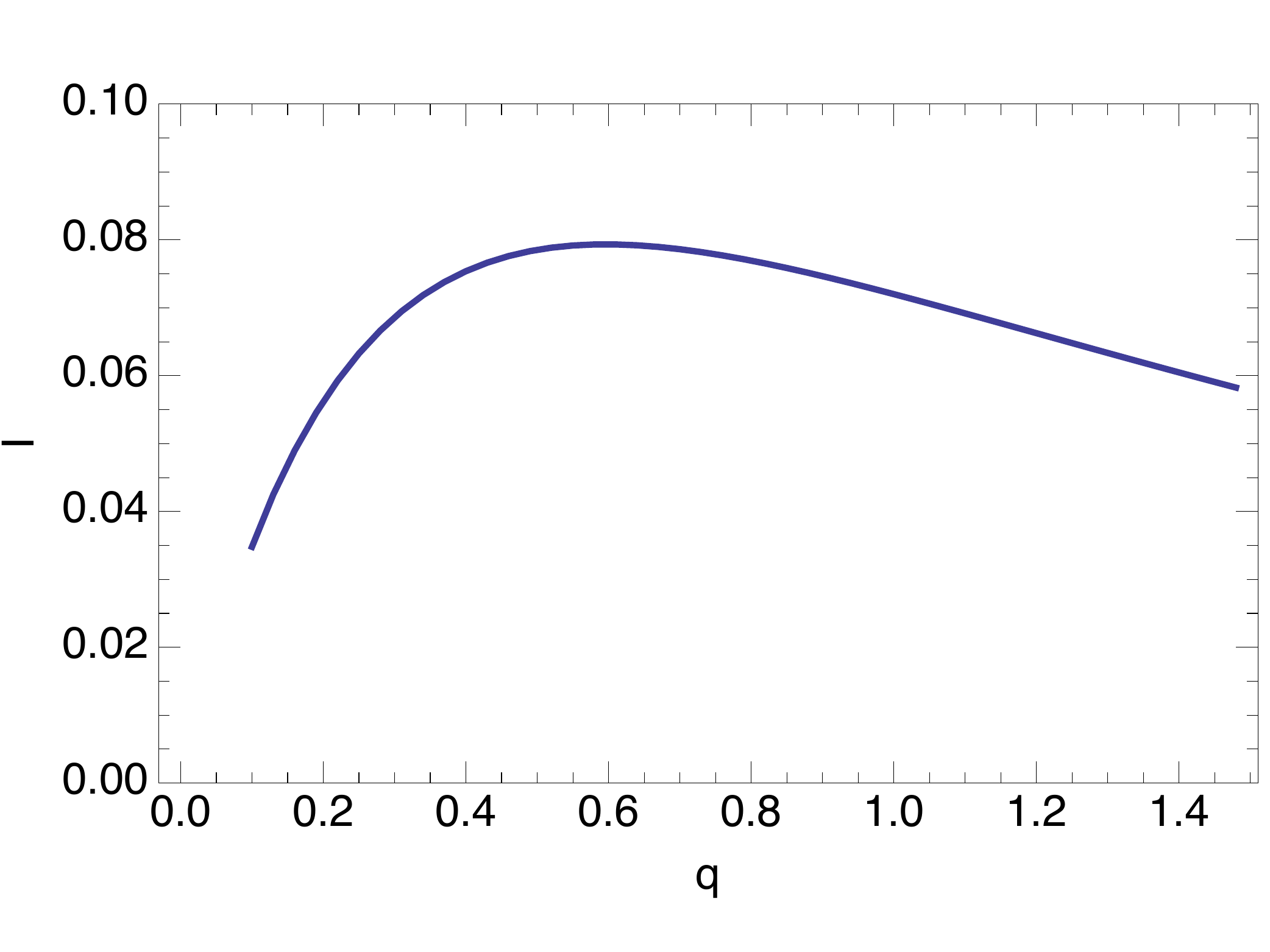}
\caption{Plot of function $I(q,q)$ versus $q$.}
\label{fig:3}
\end{figure}

We now re-write Eq~\eqref{eq:44} introducing the cooling time measured in revolution periods, $N_\mathrm{c} = t_\mathrm{c}/T$, and using the normalized variables~\eqref{eq:61}  together with $q_h\equiv {R_h\sigma_{h}\gamma}/{\Sigma}$, where $\sigma_{h}$ is the rms relative energy spread of the hadrons,
    \begin{align}\label{eq:62}
    N_\mathrm{c}^{-1}
    =
    -
    \frac{r_hc}{\pi \Sigma}
    \Re
    \int_{0}^\infty
    {d\varkappa}
    {\cal Z}(\varkappa)
    \int_{-\infty}^\infty
    \xi^2d\xi
    \left[
    i\varkappa q_h
    F_h(\xi)e^{i\varkappa q_h \xi}
    +
    F_h'(\xi)(e^{i\varkappa q_h \xi}-1)
    \right]
    .
    \end{align}
In this equation, the integration variable is $\xi=\eta/\sigma_{h}$, the distribution function $F_h$ is considered as a function of this variable (with $F_h'$ being the derivative with respect to $\xi$), and the impedance is expressed as a function of the variable $\varkappa$. Substituting the impedance~\eqref{eq:60} into Eq.~\eqref{eq:62} we find the following expression for $N_\mathrm{c}^{-1}$,
    \begin{align}\label{eq:63}
    N_\mathrm{c}^{-1}
    =
    \frac{4 I_er_hL_mL_k}{\pi \Sigma^3 \gamma^3I_A\sigma_{e}\sigma_h}
    \left[
    q_e
    \Re
    \int_{0}^\infty
    {d\varkappa}\,
    \varkappa
    e^{-\varkappa^2q_e^2/2}
    H^2
    \left(
    \varkappa
    \right)
    R(\varkappa)
    \right]
    ,
    \end{align}
where
    \begin{align}\label{eq:64}
    R(\varkappa)
    =
    i
    \sigma_h
    \int_{-\infty}^\infty
    \xi^2d\xi
    \left[
    i\varkappa q_h
    F_h(\xi)e^{i\varkappa q_h \xi}
    +
    F_h'(\xi)(e^{i\varkappa q_h \xi}-1)
    \right]
    .
    \end{align}
For a Gaussian distribution function, $F_h(\xi) = (\sqrt{2\pi}\sigma_{h})^{-1}e^{-\xi^2/2}$, the integral in Eq.~\eqref{eq:64} can be done analytically:
    \begin{align}\label{eq:65}
    R(\varkappa)
    =
	2\varkappa q_h
	e^{-\varkappa^2 q_h^2/2}
	.
    \end{align}
With this analytical expression for $R$, the expression in the square brackets in Eq.~\eqref{eq:63}, which we denote by $I$, 
    \begin{align}\label{eq:66}
    I(q_h,q_e)
    &=
    q_e
    \int_{0}^\infty
    {d\varkappa}\,
    \varkappa
    e^{-\varkappa^2q_e^2/2}
    H^2
    \left(
    \varkappa
    \right)
    R(\varkappa)
    \nonumber\\
    &=
	2 q_h
    q_e
    \int_{0}^\infty
    {d\varkappa}\,
    \varkappa^2
    e^{-\varkappa^2(q_e^2+q_h^2)/2}
    H^2
    \left(
    \varkappa
    \right)
    ,
    \end{align}
can be maximized numerically with respect to the variables $q_h$ and $q_e$ (that is the strengths of the hadron and electron chicanes in the cooling system). Note that $I$ is symmetric, $I(q_h,q_e)=I(q_e,q_h)$, hence the maximum of the integral is attained when $q_h=q_e$. The plot of function $I(q,q)$ is shown in Fig.~\ref{fig:3}; its maximum value is 0.079 at $q=0.6$.
Substituting this maximum value in Eq.~\eqref{eq:63} we arrive at the following cooling rate,
    \begin{align}\label{eq:67}
    N_{c}^{-1}
    =
    0.10\frac{1}{\gamma^3\sigma_{h}\sigma_{e}}
    \frac{I_e}{I_A}
    \frac{r_hL_mL_k }{\Sigma^3 }
    .
    \end{align}
We remind the reader that in this expression $r_h$ stands for the classical radius calculated with the charge and the mass of the hadron, $r_h = (Ze)^2/m_hc^2$. 

\begin{figure}[htb!]
\centering
\includegraphics[width=0.6\textwidth, trim=0mm 0mm 0mm 0mm, clip]{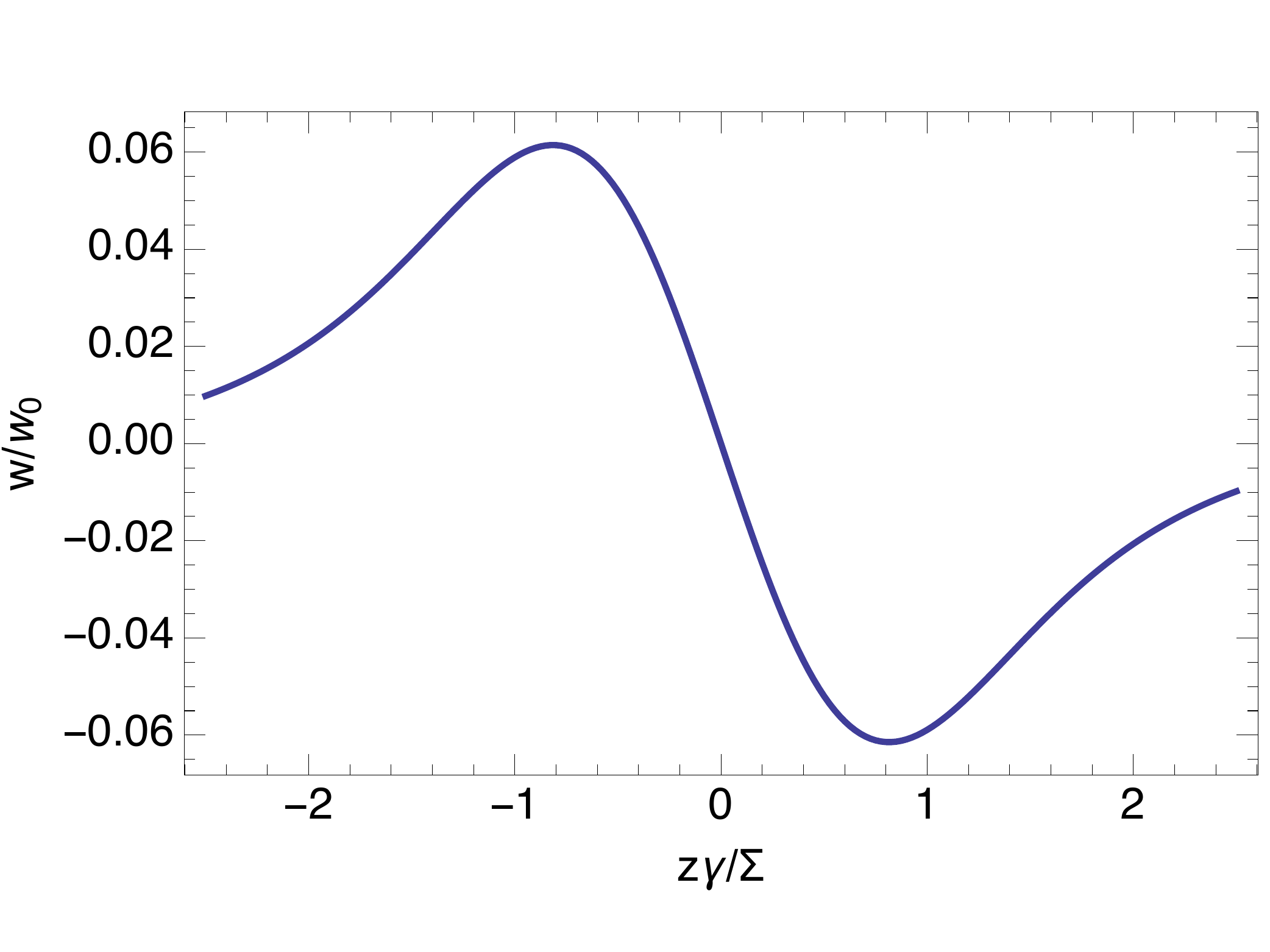}
\caption{Dimensionless wake for the hadron-electron interaction in the cooling system.}
\label{fig:4}
\end{figure}

For the optimal values of $q_h$ and $q_e$ found above, one can now calculate the interaction impedance $\cal Z$. It is more interesting, however, to find the interaction wake $w$ related to $\cal Z$ by Eq.~\eqref{eq:13}. This wake is plotted in Fig.~\ref{fig:4} as a function of the normalized variable $z\gamma/\Sigma$; the wake is normalized by the scaling factor\footnote{In the Gaussian system of units the wake has dimension of inverse length. To convert it to the SI system, one has to multiply it by $Z_0c/4\pi$.} $w_0 = 4I_eL_mL_k/\pi \Sigma^3\gamma^2I_A\sigma_e$.
This wake is an antisymmetric function of $z$ and, as has been pointed out above, is equal to zero at the origin. Numerical value of the wake for the parameters of eRHIC collider is calculated in Section~\ref{sec:9}.

%
\section{Calculation of the diffusion coefficients}\label{sec:8}
%

We can now calculate the diffusion coefficient given by Eq.~\eqref{eq:31} for the parameters of the optimal cooling. Using Eq.~\eqref{eq:60} for the impedance we find
    \begin{align}\label{eq:68}
    D
    =
    \frac{n_{0h}}{4\pi T}
    \frac{\gamma}{\Sigma}
    \left(
    \frac{4 r_hI_eL_mL_k}{\Sigma^2 \gamma^4I_A\sigma_{e}}
    \right)^2
    q_e^2
    \int_{-\infty}^\infty
    {d\varkappa}
    H^4(\varkappa)
    \varkappa^2
    e^{-\varkappa^2q_e^2}
    .
    \end{align}
Calculating the integral for the optimal value $q_e = 0.6$ we find
    \begin{align}\label{eq:69}
    q_e^2
    \int_{-\infty}^\infty
    {d\varkappa}
    H^4(\varkappa)
    \varkappa^2
    e^{-\varkappa^2q_e^2}
    =
    5.3\times10^{-3}
    .
    \end{align}
The diffusion coefficient~\eqref{eq:68} can now be written as
    \begin{align}\label{eq:70}
    D
    =
    0.66
    N_\mathrm{c}^{-2}
    \frac{I_h\Sigma}{TI_Ar_e\gamma}
    \sigma_h^2
    ,
    \end{align}
where $I_h = en_{0h}c$ is the hadron current. The requirement~\eqref{eq:34} that the diffusion does not overcome the cooling is now expressed as follows,
    \begin{align}\label{eq:71}
    0.66
    \frac{\Sigma}{\gamma r_e}
    \frac{I_h}{I_A}
    <
    1.5N_\mathrm{c}
    .
    \end{align}
In the next Section we will estimate it for the parameters of the eRHIC collider.

%
\section{Estimates for {e}RHIC collider}\label{sec:9}
%

As a numerical illustration of the general theory developed in the previous sections we will estimate the optimized cooling rate for the nominal parameters of the electron-hadron collider eRHIC~\cite{Montag:IPAC2017}. The parameters of the proton beam in eRHIC and hypothetical parameters of the electron beam in the cooling system are given in  Table~\ref{tab:1}.
\begin{table}[hbt]
\begin{center}
\begin{tabular}{lc}
\hline
\hline
Proton beam energy\hspace{70mm} 	   &  275 GeV \\
RMS length of the proton beam, $\sigma_{z}^{(h)}$  	   &  5 cm \\
RMS relative energy spread of the proton beam, $\sigma_{h}$& $4.6\times 10^{-4}$\\
Peak proton beam current, $I_h$ & 23 A\\
RMS transverse size of the beam in the cooling section, $\Sigma$ & 0.7 mm\\
Electron beam charge, $Q_e$ & 1 nC\\
RMS relative energy spread of the electron beam, $\sigma_{h}$& $1\times 10^{-4}$\\
Modulator and kicker length, $L_m$ and $L_k$  & 40 m\\
\hline
\hline
\end{tabular}
\caption{Parameters of the eRHIC collider with a hypothetical MBEC cooling section.}
\label{tab:1}
\end{center}
\end{table}

Because the cooling rate~\eqref{eq:67} depends on the local electron beam current that varies within the electron bunch, one has to average Eq.~\eqref{eq:67} taking into account the finite electron bunch length which we denote by $\sigma_{z}^{(e)}$. Assuming a Gaussian current distribution in the electron beam, $I_e = [Q_ec/\sqrt{2\pi}\sigma_{z}^{(e)}]\exp[-z^2/2(\sigma_{z}^{(e)})^2]$, it is straightforward to calculate that the average electron current that a hadron sees over many passages through the electron beam is equal to
    \begin{align}
    \bar I_e
    =
    \frac{Q_ec}{\sqrt{2\pi}[(\sigma_{z}^{(e)})^{2}+(\sigma_{z}^{(h)})^{2}]^{1/2}}
    .
    \end{align}
For an electron beam several times shorter than the hadron one, we can neglect in this formula $\sigma_{z}^{(e)}$ in comparison with $\sigma_{z}^{(h)}$. In this limit, replacing $I_e$ in Eq.~\eqref{eq:67} by $\bar I_e$, we obtain for the cooling rate
    \begin{align}\label{eq:67-1}
    N_{c}^{-1}
    =
    0.10\frac{1}{\gamma^3\sigma_{h}\sigma_{e}}
    \frac{Q_ec}{\sqrt{2\pi}\sigma_{z}^{(h)} I_A}
    \frac{r_hL_mL_k }{\Sigma^3 }
    .
    \end{align}
Substituting parameters from Table~\ref{tab:1} into this equation gives for the cooling time
    \begin{align}\label{eq:73}
    N_\mathrm{c}
    =
    1.15\times 10^{10},
    \end{align}
which, with the revolution period in the RHIC ring of $13\ \mu$s, corresponds to 41 hours. The diffusion rate estimated with Eq.~\eqref{eq:70} turns out to be much smaller than the cooling rate, so that Eq.~\eqref{eq:71} is well satisfied---the ratio of the right-hand side of Eq.~\eqref{eq:71} to its left-hand side is about $2.2\times 10^4$. The optimal parameters of the electron and proton chicanes are $R_{56}^{h}=0.31$ cm and $R_{56}^{e}=1.4$ cm. Of course, such a long cooling time is not sufficient for the eRHIC collider, where the intra-beam scattering (IBS) time scale for the emittance doubling is estimated in the range of 2 hours. We conclude that a simple setup shown in Fig.~\ref{fig:1} needs to be augmented by some kind of amplification in the electron channel, as mentioned in the Introduction. We will discuss the needed amplification factor and some of the issues related to the amplification in Section~\ref{sec:11}.

Our assumption that the hadron-electron interaction results only in the energy perturbation of electrons in the modulator, and not their density, is justified if plasma effects in the electron beam can be ignored.  Plasma oscillations convert energy perturbations in the beam into density modulations and vice versa in a quarter of the plasma wavelength $\lambda_p$, so these effects can be ignored if $\frac{1}{4} \lambda_p$ is much larger than the modulator and kicker lengths.  To estimate $\frac{1}{4} \lambda_p$ in the electron beam we can use the following formula, (see, e.g., Ref.~\cite{schneidmiller_2010}),
    \begin{align}\label{eq:72}
    \frac{1}{4}
    \lambda_p
    \sim
    \gamma^{3/2}
    \Sigma
    \sqrt{ \frac{I_A}{I_e}}
    .
    \end{align}
Substituting parameters from Table~\ref{tab:1} in this formula, and taking for the electron peak current $I_e = 30$ A we find $\frac{1}{4} \lambda_p = 84 $ m and hence $\frac{1}{4} \lambda_p \gtrsim L_m, L_k $ if the electron beam current is limited by $I_e\lesssim 30$ A.

We end this section with the calculations of the wake scaling factor $w_0$ pertinent to Fig.~\ref{fig:4}. For the eRHIC parameters, assuming the electron peak current $I_e = 30$ A, we find $\Sigma/\gamma = 2.4\ \mu$m and $w_0=1.1\times 10^{19}$ V/C. This means that the maximum/minimum values of the potential are located at $z=\pm 2\ \mu$m from the origin, and the maximum/minimum cooling potential created by a single proton in the electron beam in the kicker is $\pm 1.75$ V.

%
\section{Computer simulation of coherent cooling}\label{sec:10}
%

To test our analytical theory we carried out computer simulations of MBEC. In these simulations, electrons and hadrons are represented by macroparticles that interact with the force given by Eq.~\eqref{eq:45}. Initially, $N_e$ electron macroparticles  are randomly distributed  in the interval $0<z<\Delta z$ with the energy $\eta^{(e)}_i$ of $i$-th electron randomly assigned from a Gaussian distribution with the rms width $\sigma_e$. Periodic boundary conditions are set at the boundaries of the interval $[0,\Delta z]$. A hadron particle, with an energy $\eta^{(h)}$ randomly selected from a Gaussian distribution with the rms width $\sigma_h$, is placed at a random location within the interval and the energy of each electron $i$ is changed by $\Delta\eta_i^{(e)} = f_{z,i}L_m/\gamma m_ec^2$, where $f_{z,i}$ is the force exerting by the hadron on electron $i$. On the next step, corresponding to the passage through the chicanes, the hadron and each electron are shifted longitudinal by $R_h\eta^{(h)}$ and $R_e(\eta^{(e)}_i+\Delta\eta^{(e)}_i)$, respectively. Finally, in the kicker, the hadron energy is changed from $\eta^{(h)}$ to $\eta^{(h)}+\Delta\eta^{(h)}$ with $\Delta\eta^{(h)}=\sum_{i=1}^{N_e} f_{z,i}L_k/\gamma m_hc^2$, where now $f_{z,i}$ denotes the force acting on the hadron from $i$th electron. This procedure is repeated $M$ times and the cooling rate is estimated  as an average over $M$ runs of the difference $(\eta^{(h)}+\Delta\eta^{(h)})^2-\sigma_h^2$.

By properly scaling all dimensional variables of the simulation problem, one can find that it involves five dimensionless parameters. The first one, $\nu = n_{0e}\Sigma/\gamma$, is equal to the number of electrons on the length $\Sigma/\gamma$ and is proportional to the electron beam current. Two more parameters, $A_1$ and $A_2$, characterize the interaction strength in the modulator and the kicker normalized by the electron and hadron energy spread, respectively,
    \begin{align}\label{eq:74}
    A_1
    =
    \frac{Zr_eL_m}{\gamma\Sigma^2\sigma_e}
    ,\qquad
    A_2
    =
    \frac{r_hL_k}{Z\gamma\Sigma^2\sigma_h}
    .
    \end{align}
Finally, the last two parameters are the dimensionless strengths of the chicanes, $q_e$ and $q_h$, defined in Sections~\ref{sec:6} and~\ref{sec:7}. In the simulations we assumed $q_e=q_h=q$.

Calculating numerical values of $\nu$, $A_1$ and $A_2$ for the eRHIC parameters from Table~\ref{tab:1} and assuming the electron current $I_e = 30$ A, we find
    \begin{align}\label{eq:75}
    \nu = 1.5\times 10^6
    ,\qquad
    A_1    = 7.8\times 10^{-6}
    ,\qquad
    A_2    = 7.4\times 10^{-10}
    .
    \end{align} 
Simulations with these values are extremely difficult due to a required large number of macroparticles and small values of the interaction strengths, so we used larger values for $A_1$ and $A_2$ and a smaller value for $\nu$:
    \begin{align}\label{eq:76}
    \nu = 5\times 10^2
    ,\qquad
    A_1    = 1\times 10^{-2}
    ,\qquad
    A_2    = 9.4\times 10^{-7}
    ,
    \end{align} 
with the same ratio $A_2/A_1$ as in Eqs.~\eqref{eq:75}. Because $A_1$ and $A_2$ are proportional to the square of the charge, the increased values of $A_2$ and $A_1$ can be interpreted as if macroparticles carry a charge larger than the elementary charge $e$. Our parameter choice~\eqref{eq:76} can be interpreted as if each macroparticle has a charge of approximately $36e$.

\begin{figure}[htb]
\centering
\includegraphics[width=0.6\textwidth, trim=0mm 0mm 0mm 0mm, clip]{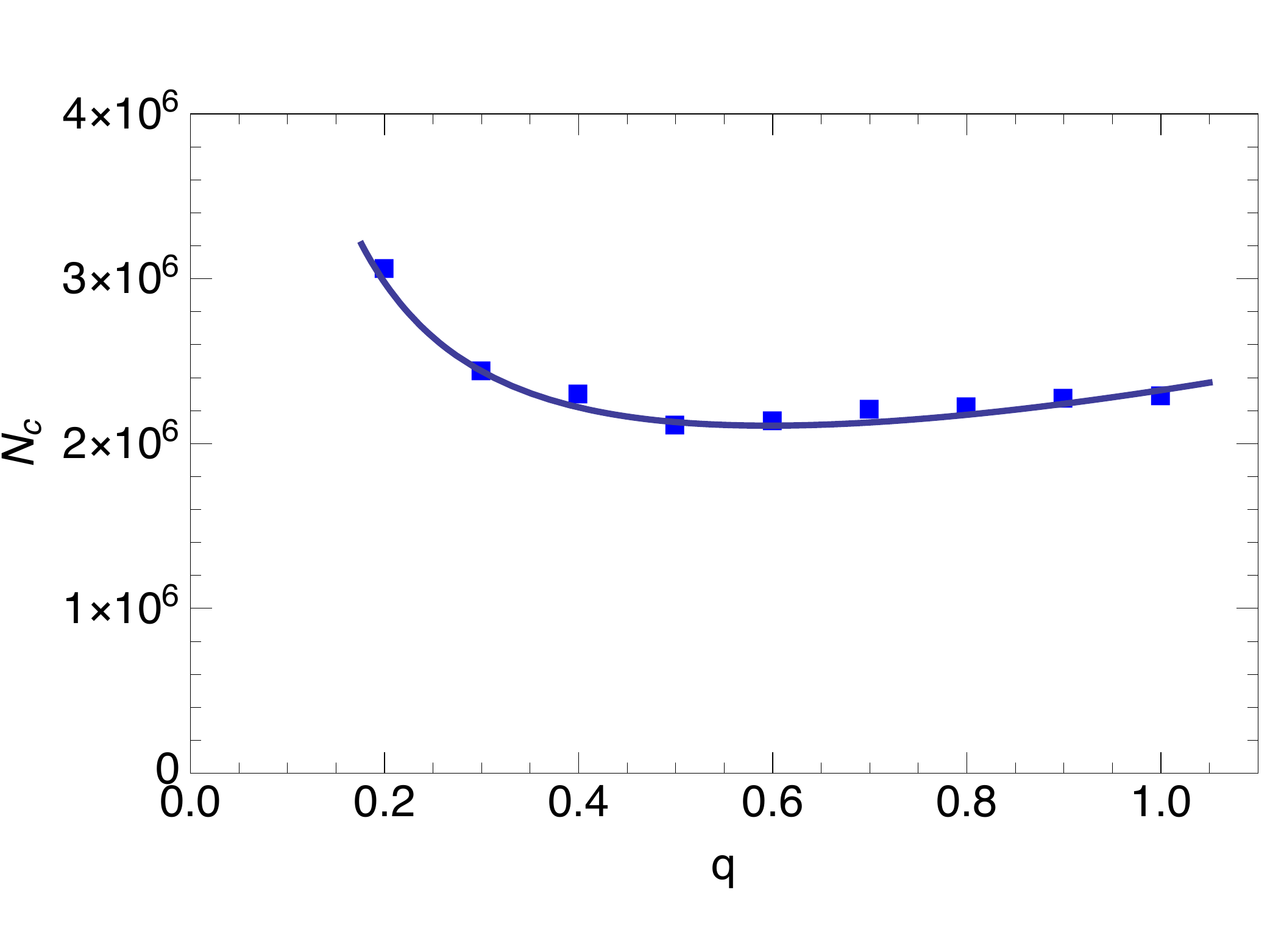}
\caption{Cooling time as a function of dimensionless chicane strength.}
\label{fig:5}
\end{figure}

We used $N_e=10^4$ electron macroparticles and the length of the ``electron bunch'' $\Delta z = 20 \Sigma/\gamma$ in the simulations. The averaging was done over $M=5\times10^6$ runs.
The plot of the simulated cooling times as a function of the dimensionless chicane strength $q$ is shown in Fig.~\ref{fig:5} by blue squares.
The solid curve is calculated using Eq.~\eqref{eq:63}. One can see that Eq.~\eqref{eq:63} is in good agreement with the simulations which we consider as a confirmation of the correctness on our analytical results.

%
\section{Cooling acceleration with amplification stages}\label{sec:11}
%

As was mentioned in Section~\ref{sec:9}, the cooling rate of the simple system shown in Fig.~\ref{fig:1} is not sufficient for the eRHIC collider without some kind of amplification system added in the electron channel. A detailed study of the specific amplification method is beyond the scope of this paper, however, based on the results of Section~\ref{sec:9} we can rather easily estimate the required gain factor that would allow to lower the cooling time below the 2 hours limit required for eRHIC. For the MBEC amplification cascades~\cite{Ratner2013} the gain factor $G$ is a broadband function of the frequency, and for a crude estimate, one can take it as a constant\footnote{In contrast to MBEC, the FEL amplification is intrinsically narrowband, and our analysis in this section is not applicable to it.}. Then $G$ appears as a multiplication factor in the expression for the impedance $\cal Z$, and the cooling rate increases by the same factor. Hence, to get the cooling time in the range of 1 hour one needs the gain factor $G\gtrsim 50$. Using the results of Ref.~\cite{schneidmiller_2010}, the amplification factor in one cascade of MBEC (consisting of a drift in which density perturbations execute one quarter of plasma oscillations followed by a chicane) can be estimated as
    \begin{align}
    G
    \sim
    \frac{1}{\sigma_e}
    \sqrt{ \frac{I_e}{\gamma I_A}}
    .
    \end{align}
From the parameters from Table~\ref{tab:1}, assuming $I_e=30$ A, we find $G\sim 24$, so we conclude that two amplification cascades should be enough to achieve the MBEC cooling time in eRHIC below one hour. A detailed theory of the MBEC cooling with amplification cascades will be published in a separate paper.

Amplification of the signal also amplifies the noise and increases the diffusion effects in the coherent cooling with the diffusion coefficient~\eqref{eq:31} scaling as $G^2$. In the inequality~\eqref{eq:71}, the left-hand side scales as $G^2$, while the right-hand side is proportional to $G$. As was mentioned in Section~\ref{sec:9}, without the amplification the left-hand side is about four orders of magnitude smaller than the right-hand side. Hence, we conclude that for $G<100$ the effect of the noise diffusion is still smaller than the cooling effect.

It is interesting to derive the maximum amplification factor, $G_\mathrm{max}$, for which the diffusion becomes of the same order as the cooling. This factor is given by the ratio of the right-hand side of Eq.~\eqref{eq:71} to the left-hand side,
    \begin{align}\label{eq:77}
    G_\mathrm{max}
    \sim
    N_\mathrm{c}
    \left(
    \frac{\Sigma}{\gamma r_e}
    \frac{I_h}{I_A}
    \right)^{-1}
    .
    \end{align}
For this maximum gain, the cooling rate becomes
    \begin{align}\label{eq:78}
    G_\mathrm{max}
    N_\mathrm{c}^{-1}
    \sim
    \frac{\gamma r_e}{\Sigma}
    \frac{I_A}{I_h}
    \sim
    \frac{\gamma n_{0h}}{\Sigma}
    .
    \end{align}
The last expression has a simple meaning---it is a number of protons in the amplification bandwidth $\Sigma/\gamma$---in agreement with the general principles of the stochastic cooling~\cite{mohl2013stochastic}. For the parameters from Table~\ref{tab:1}, this bandwidth is estimated as $c\gamma/\Sigma \approx 2\pi\times 20$ THz, and is much larger than the typical bandwidth of several GigaHertz in a typical classical stochastic cooling setup.

%
\section{Discussion}\label{sec:12}
%

In this paper, we derived the cooling rate for the longitudinal, or momentum, cooling using a simple 1D model that treats particles as charged disks interacting through the Coulomb force. There are several effects that are neglected in this model. Clearly, the transverse dynamics due to the beam focusing is ignored, as well as longitudinal displacement of particles due to this focusing. We also ignored plasma oscillations in the electron beam in the modulator and the kicker regions. This is justified if the length of the modulator and the kicker is smaller than a quarter of the plasma period in the electron beam. As was estimated in Section~\ref{sec:9}, this requirement is satisfied for the parameters of a MBEC cooler for eRHIC. 

In our analysis, we assumed a round cross section of the beams with a Gaussian radial density distribution. This assumption can be easily dropped and other transverse distributions (e.g., with unequal vertical and horizontal sizes) used for the particle interaction. This will only change the specific form of the interaction potential~\eqref{eq:46}, with the rest of the calculations of the cooling rate remaining the same. 

Finally, we note that the 1D theory can also be extended to include the effects of the transverse cooling. This type of cooling is achieved through the introduction of the dispersion in the modulator and the kicker regions, as it was proposed for the optical stochastic cooling scheme~\cite{OSC_RHIC,Lebedev:2014cha}. A preliminary consideration of the horizontal emittance cooling in MBEC has been carried out in Ref.~\cite{baxevanis18_s}.

%
\section{Acknowledgements}\label{sec:13}
%

I would like to thank M. Blaskiewicz, F. Willeke and M. Zolotorev for numerous stimulating discussions of the subject of this paper. I am also grateful to E. Shneidmiller and M. Dohlus for clarifying the connection of MBEC with microbunching instability in FELs, and to P. Baxevanis for useful comments and help with computer simulations.

This work was supported by the Department of Energy, contract DE-AC03-76SF00515.
\appendix
%
\section{Derivation based on analysis of particle-to-particle interactions}\label{sec:app1}
%

The averaged energy loss~\eqref{eq:28} and the diffusion coefficient~\eqref{eq:31} can also be obtained from a straightforward consideration of particle interactions through the wakefield. The relative energy change $\Delta \eta_{i}$ of a particle $i$ due to such interaction is
    \begin{align}\label{eq:A.1}
    \Delta \eta_{i}
    &=
    \frac{1}{E_0}
    \Delta E_{j}
    =
    \frac{e^{2}}{E_0}
    \left[
    w(0)
    +
    \sum_{l\ne i}
    w(z_{i}-z_{l})
    \right]
    ,
    \end{align}
where we have included the term $w(0)$ responsible for the interaction of the particle with itself. Calculating the averaged value $\langle\Delta \eta_{i}\rangle$ one has to take into account that the average value of the sum on the right-hand side of Eq.~\eqref{eq:A.1} is equal to zero, because in an infinitely long uniform bunch the averaging can be replaced by the integration over $z_i$ and the wake function has a zero average,
    \begin{align}\label{eq:A.2}
    \intinf
    dz
    w(z)
    =
    0
    .
    \end{align}
Hence $\langle\Delta \eta_{i}\rangle = {e^{2}}w(0)/{E_0}$ which is equivalent to say that the average energy loss is given by Eq.~\eqref{eq:28}. 

The diffusion coefficient~\eqref{eq:31} can be expressed through the averaged square of the energy deviation in one step:
    \begin{align}\label{eq:A.3}
    D
    =
    \frac{1}{2T}
    \langle
    (\Delta \eta - \langle\Delta \eta\rangle)^2
    \rangle
    =
    \frac{1}{2T}
    [\langle(\Delta \eta)^2\rangle - \langle\Delta \eta\rangle^2]
    .
    \end{align}
For the averaged square we have
    \begin{align}\label{eq:A.4}
    \langle
    \Delta \eta_{j}^{2}
    \rangle
    &=
    \frac{e^{4}}{E_0^{2}}
    \left[
    w(0)
    +
    \sum_{l\ne j}
    w(z_{j}-z_{l})
    \right]
    \left[
    w(0)
    +
    \sum_{m\ne j}
    w(z_{j}-z_{m})
    \right]
    \nonumber\\
    &=
    \langle \Delta \eta \rangle^{2}
    +
    \frac{e^{4}}{E_0^{2}}
    \sum_{l,m\ne j}
    w(z_{j}-z_{l})
    w(z_{j}-z_{m})
    .
    \end{align}
In the last term on the right-hand side we have both 2-particle (when $m=l$) and 3-particle  (when $m\ne l$) interactions. The non-zero contribution comes from the 2-particle interactions only, which can be expressed through the impedance,
    \begin{align}\label{eq:A.5}
    \langle(\Delta \eta)^2\rangle - \langle\Delta \eta\rangle^2
    =
    \frac{e^{4}}{E_0^{2}}
    \sum_{m\ne j}
    w(z_{j}-z_{m})^{2}
    &
    \to
    \frac{e^{4}}{E_0^{2}}
    n_{0}  
    \int_{-\infty}^{\infty}
    dz  
    w(z)^{2}
    \nonumber\\&
    =
    \frac{e^{4}}{E_0^{2}}
    \left(
    \frac{c}{2\pi}
    \right)^{2}
    n_{0}  
    \int_{-\infty}^{\infty}
    ds  
    \int_{-\infty}^\infty
    dk
    dk'
    {\cal Z}(k)
    {\cal Z}(k')
    e^{-i(k+k')s}
    \nonumber\\&
    =
    \frac{e^{4}c^{2}}{2\pi E_0^{2}}
    n_{0}  
    \int_{-\infty}^\infty
    dk
    |{\cal Z}(k)|^{2}
    .
    \end{align}
Substituting this term to Eq.~\eqref{eq:A.5} gives the diffusion coefficient~\eqref{eq:31}.

\bibliography{\string~/gsfiles/Bibliography/master%
              }
\end{document}